\documentclass[aps, prb, superscriptaddress, showpacs, showkeys, twocolumn, footinbib]{revtex4-1}
\usepackage{graphicx}
\usepackage{physics}

\usepackage{dsfont} 
\usepackage{bm}

\usepackage{hyperref}
\usepackage{todonotes}

\newcommand{\ie}{\emph{i.e.}~}
\newcommand{\via}{\emph{via}~}

\newcommand{\cf}{\emph{cf.}~}
\newcommand{\eg}{\emph{e.g.,}~}
\graphicspath{{Figures}{./Figures/}{../Figures/}}
\newcommand{\hc}{{\rm h. \, c.}}
\newcommand{\id}{\mathds{1}}

\newcommand\Eq[1]{Eq.~(\ref{#1})}
\newcommand\Fig[1]{Fig.~\ref{#1}}

\newcommand{\Do}{\Delta_0}

\usepackage{amsmath}
\usepackage{amsfonts}
\usepackage{amssymb}
\usepackage{amsthm}
\usepackage{lipsum}
\renewcommand{\H}{{\cal H}}
\newcommand{\up}{\uparrow}
\newcommand{\down}{\downarrow}
\newcommand{\Rsk}{R_{\rm sk}}
\newcommand{\neel}{N\'{e}el~}

\begin{document}

\title{Topological superconductivity with orbital effects in magnetic skyrmion based heterostructures}

\author{Maxime Garnier}
\email[Corresponding author: \,]{maxime.garnier1@u-psud.fr}
\affiliation{Laboratoire de Physique des Solides, UMR 8502, CNRS,
Universit\'{e} Paris-Sud, Universit\'{e} Paris-Saclay, 91405 Orsay, France}
\author{Andrej Mesaros}
\affiliation{Laboratoire de Physique des Solides, UMR 8502, CNRS,
Universit\'{e} Paris-Sud, Universit\'{e} Paris-Saclay, 91405 Orsay, France}
\author{Pascal Simon}
\affiliation{Laboratoire de Physique des Solides, UMR 8502, CNRS,
Universit\'{e} Paris-Sud, Universit\'{e} Paris-Saclay, 91405 Orsay, France}

\date{\today}

\begin{abstract}
Proximitizing magnetic textures and $s$-wave superconductors is becoming a platform for engineering topological superconductivity and Majorana fermions by the means of exchange processes. However, the consequences of orbital effects have not yet been fully taken into account. In this work, we investigate the magnetic skyrmion texture-induced orbital effects using a Ginzburg-Landau approach and clarify the conditions under which they can induce superconducting vortices. These orbital effects are then included in Bogoliubov-De-Gennes theory containing the exchange interaction, as well as superconducting vortices (when induced). We find that the topological phase is largely stable to all investigated effects, increasing the realistic promise of skyrmion-superconductor hybrid structures for realization of topological superconductivity.
\end{abstract}
\maketitle

\section{Introduction}

The search for topological superconductors and  Majorana fermions has mostly relied on an approach combining conventional $s$-wave superconductivity, spin-orbit coupling and magnetism to generate $p$-wave pairing. Even though great advances have been made using this approach using semiconducting wires 
(see \eg Ref.~\onlinecite{Lutchyn:2018} for a review) or magnetic atoms combined with 
superconducting substrates,\cite{NP2014, FrankePRL:2015, Pawlak2016, Wiesendanger_Chain_2018, MenardPbCoSi, Menard:2019vn, Palacio-Morales:2019} these experiments remain highly challenging so that it may be worth considering removing one of the ingredients. For example, one may consider removing the spin-orbit coupling since the exchange interaction between conduction electrons and a magnetic texture induces an effective spin-orbit interaction.\cite{Braunecker2010,Choy2011,NP2013, Braunecker2013, Klinovaja2013, Vazifeh2013, Pientka2013} This seems to be a viable platform for inducing topological superconductivity.\cite{Choy2011,Nakosai2013, Schnyder2015, Zutic_PRL, Mohanta2019, Kontos2019} In particular, particle-like topological spin configurations known as magnetic skyrmions\cite{TokNag2013, Fert2013, Everschor-Sitte:2018} have recently been highlighted as prime candidates\cite{Loss_Majorana_skyrmion, KovalevPRB2018, Morr_engineering, Garnier_SkTopoSC, Rex_SK_Vortex_2019} with interesting prospects for manipulation due to their high degree of controllability.\cite{Rudner2016, Milosevic2019} However, in this context, the magnetic orbital effects have not yet been fully taken into account. 

Generally, systems in presence of supercurrents and spin-orbit coupling may support topologically non-trivial phases without a spin coupling of the Zeeman form.\cite{Romito:2012, Kotetes:2015, fullshell_theo_2018, fullshell_exp_2018, Akhmerov_2019} This suggests that orbital effects are sufficient to create topological phases in magnet-superconductor heterostructures. In a more involved scenario, the Zeeman-form coupling (induced by the exchange interaction between skyrmion and superconductor) combined with the magneto-electric effect due to the spin-orbit coupling leads to the appearance of vortices,\cite{Rudner2016,Baumard_Buzdin_2019} which induce supercurrents and topological superconductivity.\cite{Rex_SK_Vortex_2019} If one now considers alternatives based on removing the explicit spin-orbit coupling, which is a key coupling ingredient in previous scenarios, one finds that topological superconductivity is also predicted with {\it only} the Zeeman-form coupling induced by the exchange interaction with the skyrmion.\cite{Loss_Majorana_skyrmion, KovalevPRB2018, Morr_engineering, Garnier_SkTopoSC, Rex_SK_Vortex_2019} However, even without explicit spin-orbit coupling, one should consider the effect of supercurrents, which may appear only \via the orbital effects generated by the skyrmion\cite{Eremin_2019} (since the magneto-electric coupling vanishes together with the spin-orbit coupling). The orbital effects due to skyrmions cannot be generally neglected, as we argue below, and their effect on topological superconductivity has not been analyzed so far. As a consequence, it is not clear if vortices are to be expected and whether topological superconductivity persists when both exchange and orbital effects of the skyrmion are included in a superconductor without spin-orbit coupling.

In conventional type-II superconducting thin films, the effective penetration depth (or Pearl length) $\lambda_{\rm eff}$ can be orders of magnitude larger than the film thickness $d$.\cite{Pearl:1964, deGennes_SC, Gubin:2005} This implies that the screening currents are weak and may become negligible. Previous works\cite{Rex_SK_Vortex_2019,Rudner2016} have used this observation to set to zero the magnetic vector potential in the superconductor. However, we argue that since the screening is weak, the magnetic field generated by the skyrmion penetrates the superconductor almost unaltered, and is thus not necessarily negligible. Therefore in contrast to previous works, we include the vector potential generated by the skyrmion as an orbital effect on the electrons in the superconductor. To fully understand the phase diagram, we consider the magnetic exchange as an independent effect on the electrons, since this term could be experimentally tuned by an insulating non-magnetic layer, which prevents the hopping of electrons between the skyrmion and the superconductor while not affecting the strength of the vector potential.

The aim of the present work is twofold. First, we investigate the orbital effects of the magnetic field created by the skyrmion on a conventional type-II superconductor without spin-orbit coupling and clarify the conditions of existence of vortices in a Ginzburg-Landau framework. Second, using the Bogoliubov-de-Gennes formalism we address the robustness of the topological phase induced by the exchange interaction to the orbital effects and the possible presence of vortices. We find that the exchange-induced topological phase is robust to the inclusion of orbital effects, even in the presence of vortices. This contributes to making the skyrmion-superconductor heterostructure an even more promising platform for the realization of topological superconductivity.

The article is organized as follows. In Sec.~\ref{sec: setup}, we present the system under study, the hypotheses made on the superconductor, derive an expression for the magnetic field generated by a single skyrmion and set up our Ginzburg-Landau analysis. We establish the phase diagram of the superconductor in Sec.~\ref{sec: PhDiag} and study the implications on the topological superconducting phase in Sec.~\ref{sec: TSC}.

\section{Setup}
\label{sec: setup}

\begin{figure}
\centering
\includegraphics[width= \columnwidth]{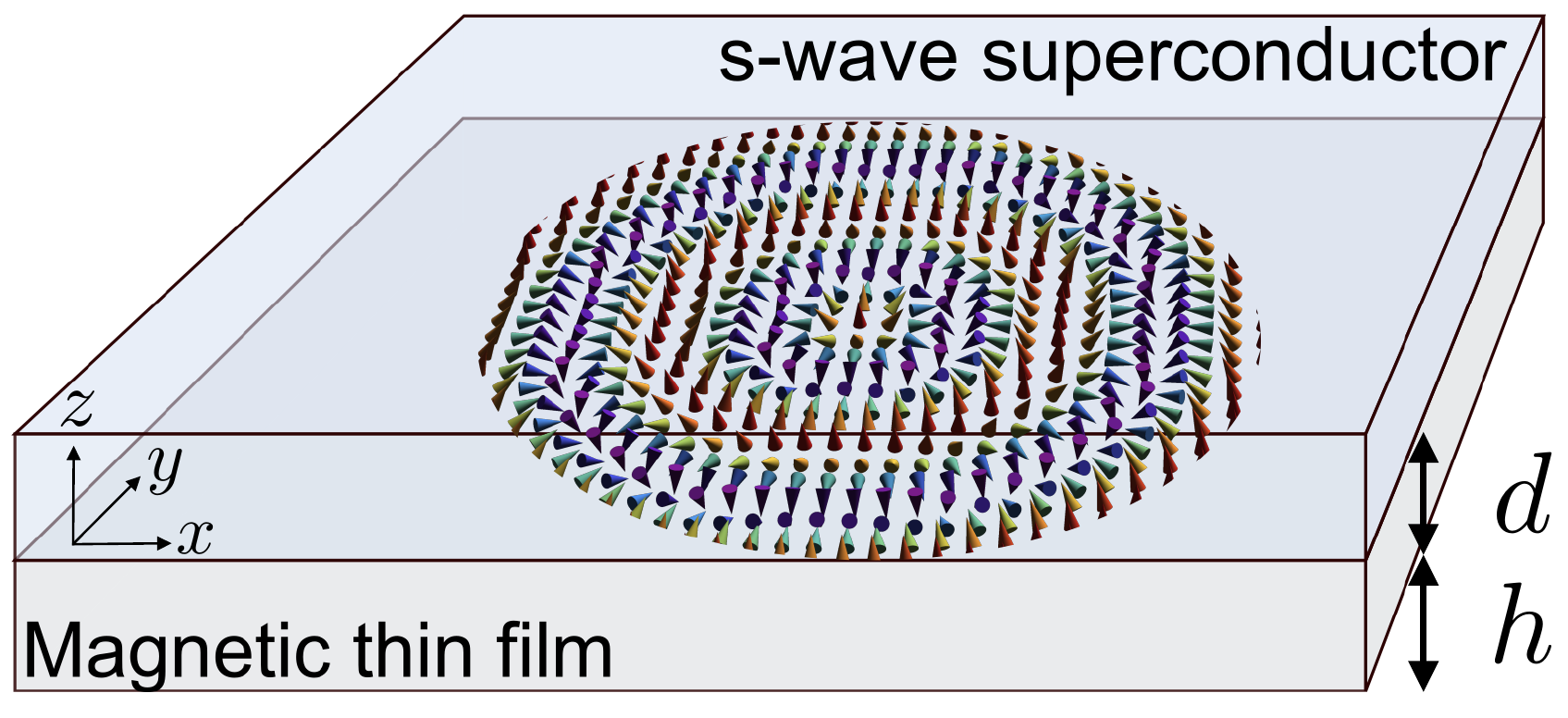}
\caption{Schematics of the system. A conventional superconductor of thickness $d$ in proximity to an insulating magnetic film of thickness $h$ harboring a \neel skyrmion with radial winding $p = 4$ (see \Eq{eq: skyrmion_def}). The arrows represent the magnetization of the skyrmion and are color-coded according to their $z$-axis projection.}
\label{fig: setup}
\end{figure}

The system under investigation is composed of a conventional type-II  superconducting film without spin-orbit coupling (such as Al) in proximity to an insulating magnetic thin film harboring a skyrmion (such as ${\rm Cu}_2{\rm O}{\rm Se}{\rm O}_3$) \cite{Everschor-Sitte:2018,Seki2012} as shown in \Fig{fig: setup}. We denote by $d$ the thickness of the superconducting layer and by $h$ the thickness of the magnet.
We focus on insulating magnets so as to minimize feedback mechanisms on the magnet due to the superconductor. \cite{Hals-SOT:2016, Motome2018, NogueraPRB2018} Even though magnetic skyrmions exist in two types, namely \neel (\cf \Fig{fig: setup}) and Bloch (as in ${\rm Cu}_2{\rm O}{\rm Se}{\rm O}_3$),\cite{TokNag2013, Fert2013} this distinction does not modify the physics described here as will be clarified.

Since in-plane critical fields of superconducting films are usually larger than that of bulk superconductors\cite{Tinkham_SC}, in analyzing the magnetostatics of our system we focus on the magnetic field component perpendicular to the superconductor ($z$ axis). The orbital coupling to the superconductor is hence through the vector potential which satisfies ${\bm \nabla} \cross {\bf  A}\left({\bf r}, z\right) = B_z\left({\bf r}\right)\bf{\hat{z}}$, where ${\bf r}$ labels the position in the plane. We take $B_z$ to be the magnetic induction created by a lone skyrmion, thereby neglecting any feedback effect of the superconductor on the magnetic material. This assumption is consistent with neglecting the screening currents and their fields. In the limit where the skyrmion is confined to a plane, the magnetic induction $B_z\left({\bf r}, z\right)$ it creates becomes equal to its magnetization component $m_z\left({\bf r}\right)$ at least near the plane (see Ref.~\onlinecite{QinWang_SKL_2018} and Supplementary Material (SM)~\ref{sec: app_SkL}). More precisely, we find that  $B_z$ decays away from the plane on a lengthscale given by the radius $\Rsk$ of the skyrmion, and for $z$ smaller than this lengthscale we can define ${\bf  A}$ by ${\bm \nabla} \cross {\bf  A} = \mu_0 \, m_z\left({\bf r}\right)\bf{\hat{z}}$ where $\mu_0$ is the vacuum magnetic permeability. The radius $\Rsk$ is defined by the magnetization profile ${\bf m}\left({\bf r}\right)$ of the skyrmion texture in polar coordinates ${\bf r} = (r, \theta)$ 
\begin{align}
{\bf m}\left({\bf r}\right)  =  M
\begin{pmatrix}
\sin f(r) \cos \left(\theta + \gamma\right)\\
\sin f(r) \sin \left(\theta + \gamma\right)\\ 
\cos f(r)
\end{pmatrix}
\label{eq: skyrmion_def}
\end{align}
The skyrmion is characterized by the radial winding number $p \in {\mathbb N}$ that counts the number of spin flips as one moves away from the core $r = 0$ along the radial direction. $f(r)$ is the radial profile of the skyrmion that we choose to be $f\left(r\right) = k_s r$ for $r \leq \Rsk$, where we have introduced $k_s = \pi/\lambda_s$ with the spin-flip length $\lambda_s = \Rsk/p$. The global angular offset $\gamma$ called helicity\cite{TokNag2013} allows to describe both \neel ($\gamma$ = 0, $\pi$) and Bloch ($\gamma$ = $\pm \frac{\pi}{2}$) skyrmions. The norm $M$ of ${\bf m}\left({\bf r}\right)$ defines the saturation magnetization of the magnet.

Focusing on the \neel ($\gamma = 0$) case, the magnetization \Eq{eq: skyrmion_def} can be written in cylindrical coordinates
\begin{align}
{\bf m}\left({\bf r}\right) = M \left(\sin\left(k_s r\right) {\bf u}_r + \cos\left(k_s r\right){\bf u}_z \right) 
\end{align}
A suitable vector potential is
\begin{align}
{\bf A}\left({\bf r}\right) =
\frac{\mu_0 M}{k_s}\left(\dfrac{\cos \left(k_s r\right)}{k_s r} + \sin \left(k_s r\right)\right) \,{\bf u}_\theta
\label{eq: SK_vect_pot}
\end{align}% Coulomb gauge is ok.
which indeed gives $B_z = \mu_0 M \cos \left(k_s r\right)$. Note that although this expression was derived for a \neel skyrmion, the magnetic induction generated by a Bloch skyrmion is qualitatively the same (see Ref.~\onlinecite{QinWang_SKL_2018} and SM~\ref{sec: app_SkL}).

For conventional superconductors, the Ginzburg-Landau (GL) free energy functional ${\cal F}$ is
\begin{align}
{\cal F} = \frac{B_c^2}{\mu_0}\int d^3{\bf r} \left[-\left\vert{\tilde \Delta}\right\vert^2 + \frac{1}{2}\left\vert{\tilde \Delta}\right\vert^4 + \xi^2\left\vert\hat{{\bf D}}{\tilde \Delta}\right\vert^2 +\dfrac{{\bf B}^2}{2B_c^2}\right]
\label{eq: GL_F}
\end{align}
where ${\tilde \Delta}$ is the superconducting order parameter normalized by its thermodynamic value in the absence of fields or gradients, $B_c$ is the thermodynamic critical field, $\hat{{\bf D}} = -i{\bm \nabla} +\frac{2e}{\hbar}\bf{A}$ is the covariant derivative with $2e$ the charge of Cooper pairs, and $\xi$ is the superconducting coherence length.\cite{Tinkham_SC}
Finally, ${\bf B} = {\bm \nabla} \cross {\bf A}$ is the magnetic induction inside the superconductor.  The free energy \Eq{eq: GL_F} is measured with respect to the normal-state free energy. As we are interested in the behavior of the superconductor in an external magnetic field, the correct thermodynamic potential to consider is the Gibbs free energy ${\cal G} = {\cal F} - \int d^3{\bf r} \, {\bf H}\left({\bf r}\right)\cdot{\bf B}\left({\bf r}\right)$ where ${\bf H}$ is the magnetic field.
As mentioned above, the typical electromagnetic response of a type-II superconducting film in a \textit{homogeneous} magnetic field occurs on a lengthscale $\lambda_{\rm eff}$ much larger than any other lengthscale in the problem. As a consequence, the {\it inhomogeneous} response on lengthscales of the order of the skyrmion's spin-flip length $\lambda_s$ cannot be inferred easily. As outlined above we neglect the screening currents so that we can approximate ${\bf B}$ by $\mu_0{\bf H}$ inside the superconductor.\cite{FetterInParks}
With this assumption, ${\cal G}$ simply reduces to ${\cal F}$ and is given by \Eq{eq: GL_F} where ${\bf B}$ is now the skyrmion-generated induction given below \Eq{eq: SK_vect_pot}.

We further suppose that all quantities are independent on the $z$ coordinate, which is valid as long as the superconducting thin film thickness $d$ is smaller than the skyrmion radius $\Rsk$, so that in this regime the thickness $d$ can be factored out of the free energy.

The final ingredient of our model is a superconducting vortex. To establish its presence or absence we use a standard vortex ansatz\cite{Tinkham_SC}  ${\tilde \Delta}\left({\bf r}\right) = {\tilde \Delta_\alpha}\left(r\right) \,e^{i \alpha \theta}$ where ${\tilde \Delta_\alpha}\left(r\right) =\tanh^{\left\vert\alpha \right\vert} \left(r/\xi\right) $ and $\alpha \in \mathbb{Z} $ is the phase winding of the vortex. Note that even if this ansatz corresponds to an Abrikosov vortex (\ie in  a bulk sample), we expect that the details of the ansatz don't matter much as long as the order parameter amplitude decays on a lengthscale $\xi$ and vanishes at the vortex core. 
Including the vector potential due to the skyrmion \Eq{eq: SK_vect_pot} and the vortex ansatz, the total free energy ${\cal G}_{\rm tot}$ is:
\begin{widetext}
\begin{align}
  {\cal G}_{\rm tot} = \frac{B_c^2}{\mu_0}\frac{2\pi d}{k_s^2} \int \!\textrm{d}{\tilde r}\, \tilde{r} \left[-\left\vert{\tilde \Delta_\alpha}\right\vert^2 + \frac{1}{2}\left\vert{\tilde \Delta_\alpha}\right\vert^4 +  k_s^2 \xi^2 \left(\partial_{\tilde{r}}\left\vert{\tilde \Delta_\alpha}\right\vert\right)^2  + \frac{k_s^2 \xi^2\nu^2}{ \tilde{r}^2}\left\vert{\tilde \Delta_\alpha}\right\vert^2\left(\frac{\alpha}{\nu} + \cos\left(\tilde{r}\right) + \tilde{r}\sin\left(\tilde{r}\right)\right)^2 +\frac{p^2\pi^2}{8}\left(\frac{\mu_0 M}{B_c}\right)^2
    %\dfrac{{\bf B}^2}{2B_c^2}
  \right] 
\label{eq: adim_G}
\end{align}
\end{widetext}
where we have defined the dimensionless coordinate ${\tilde r} = k_s r$ and the parameter $\nu \equiv \frac{2\pi \mu_0 M}{k_s^2 \phi_0}$ where $\phi_0 = h/2e$ is the superconducting flux quantum. The $\nu$ parameter can be interpreted as the ratio $ \left\vert{\phi_S}\right\vert / \left(2\phi_0\right)$ of the flux of the skyrmion $\left\vert{\phi_S}\right\vert$ through a disc of radius $\lambda_s$ (the single spin flip distance) to the normal-metal flux quantum $2\phi_0$.

\section{Phase diagram of the superconductor without exchange effects}
\label{sec: PhDiag}

The free energy, \Eq{eq: adim_G}, depends on three parameters:
\begin{itemize}
\item[({\it i})] the ratio of the characteristic skyrmion flux and the flux quantum $\nu = \frac{\left\vert\phi_S\right\vert}{2 \phi_0}$,
\item[({\it ii})] the ratio of the superconducting coherence length to the spin-flip length of the skyrmion $k_s \xi=p \pi \xi/\Rsk$,
\item[({\it iii})] the ratio $\mu_0 M / B_c$.
\end{itemize}
Given some values of these parameters, our strategy is to numerically compute the free energy for different windings $\alpha$ (including $\alpha=0$ for absence of vortex) and find the value that gives the lowest free energy. If the free energy is negative, the normal state is realized instead of the superconducting one.
Furthermore, as the magnetic energy term (last term in \Eq{eq: adim_G}) doesn't depend on $\alpha$, we start by neglecting it.

\begin{figure}
\centering
%\hspace*{-.5cm}
\includegraphics[scale = .4]{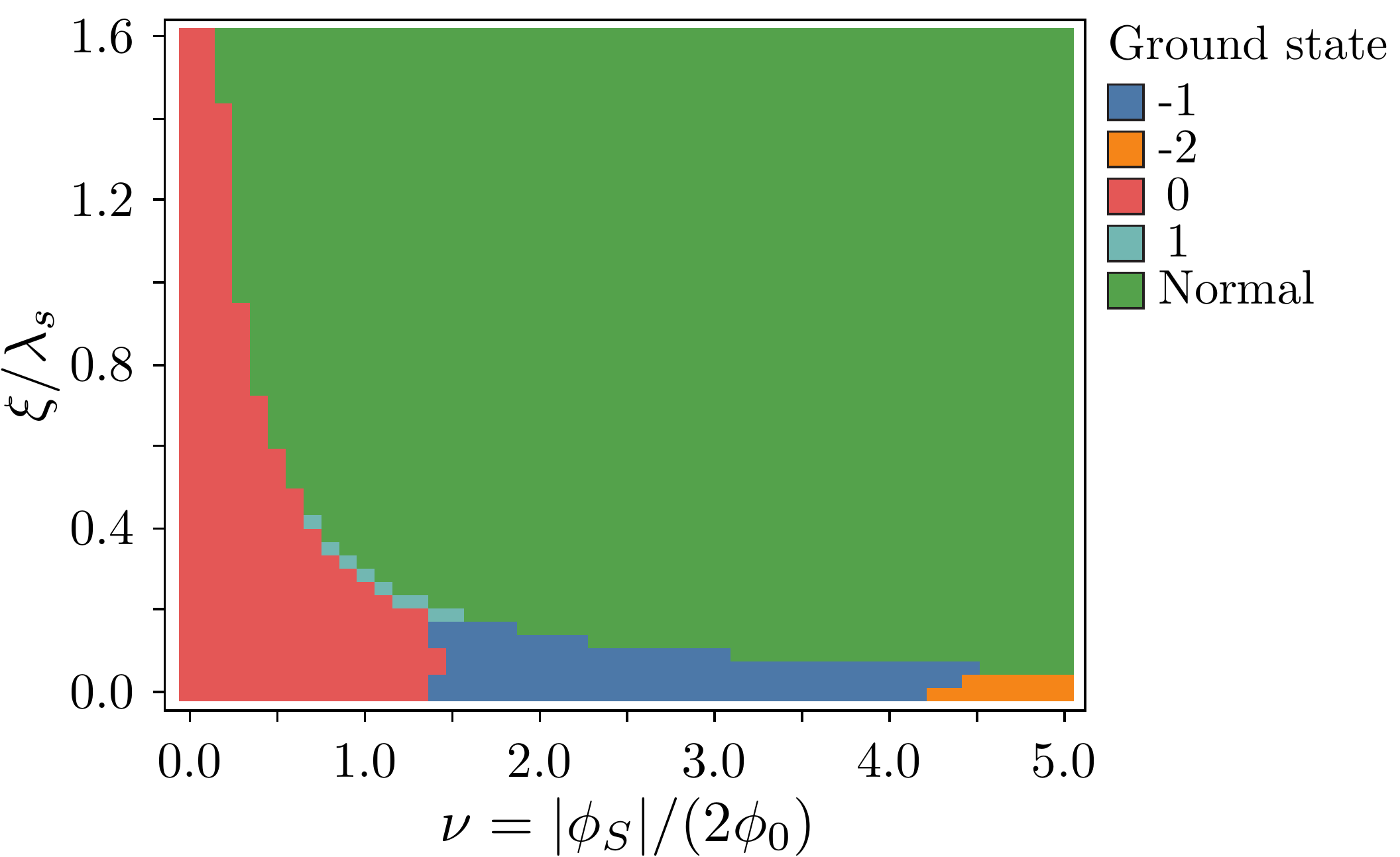}
\caption{Phase diagram obtained from \Eq{eq: adim_G} for a $p = 4$ skyrmion with short-distance cutoff $k_s l = 10^{-1}$. If the ground state is superconducting, the color represents the value of the vortex winding number $\alpha$, with $\alpha=0$ meaning absence of vortex (red). The non-superconducting ground state is labeled ``normal'' (green).}
\label{fig: PhDiag}
\end{figure}

In \Fig{fig: PhDiag} we show the phase diagram obtained for a $p=4$ skyrmion  where we have introduced a short-distance cutoff $k_s l$ for the dimensionless variable ${\tilde r}$ to deal with the logarithmic divergence in the $\alpha = 0$ case. (This cutoff has only minor consequences on our results, see SM~\ref{app: cutoff}, and its value is chosen to be comparable to the lattice spacing in the Bogoliubov-de-Gennes Hamiltonian of Sec.~\ref{sec: TSC}).

The phase diagram shows that there exists a superconducting phase without vortex (red color) for a relatively large range of superconducting coherence lengths and skyrmion fluxes $\nu \lesssim 1.5$. For sufficiently strong skyrmion flux and small $\xi/\Rsk$, there exist phases with a vortex.
We have also checked that including the magnetic energy term and varying the size of the skyrmion don't affect qualitatively our results, see SM~\ref{app: magterm} and \ref{app: SkSize}. 
We thus conclude that vortices are not always expected when proximitizing a conventional superconductor without spin-orbit coupling by a magnetic skyrmion.

\section{Implications for topological superconductivity}
\label{sec: TSC}

We now turn to the consequences of the orbital effects on the topological superconducting phase induced by the exchange interaction.\cite{Loss_Majorana_skyrmion, Garnier_SkTopoSC, Rex_SK_Vortex_2019} As shown in the previous section, the inclusion of the orbital effects due to the skyrmion has two consequences: the first is that electrons see a magnetic vector potential and the second is that vortices may be present.

\subsection{Bogoliubov-de-Gennes setup}

In the Bogoliubov-de-Gennes (BdG) formalism, the total Hamiltonian $H$ describing the electrons can be written $H = \frac{1}{2}\int d{\bf r} \, \Psi^\dagger \left({\bf r}\right)  \H\left({\bf r}\right)  \Psi \left({\bf r}\right) $ where $\H$ is the BdG Hamiltonian. Throughout the article, we work in the Nambu basis $\Psi^\dagger \left({\bf r}\right) = \left(\psi_\up^\dagger\left({\bf r}\right) , \psi_\down^\dagger\left({\bf r}\right),   \psi_\down\left({\bf r}\right) , -\psi_\up\left({\bf r}\right) \right)$ where $\psi_\sigma^\dagger\left({\bf r}\right) $ is the field operator creating an electron with spin projection $\sigma = \up,\, \down$ at position ${\bf r} = \left(r, \theta\right)$ in two dimensions (2D). 
Following the minimal coupling procedure $\hat{{\bf p}} \rightarrow \hat{{\bf p}} +e {\bf A}$, the 2D BdG Hamiltonian in the presence of both orbital and exchange effects reads
\begin{align}
\H\left({\bf r}\right)  = \left(\frac{1}{2m}\left(\hat{\bf{p}} + e{\bf A} \tau_z\right)^2 - \mu\right) \tau_z + J	\, \bm{\sigma} \cdot {\bf m}\left({\bf r}\right)  + \Delta_0 \, \tau_x
\label{eq: BdG_Skyrmion_full}
\end{align}
where $\tau_i$ and  $\sigma_i$, $i = x,\, y,\, z$, are Pauli matrices acting in particle-hole and spin space, respectively, $\hat{\bf{p}} = -i\hbar {\bm \nabla}$ is the momentum operator, $m$ is the effective mass, $\mu$ the chemical potential, $\Do$ the bare $s$-wave pairing, $J$ is the exchange interaction, and orbital effects are due to the vector potential {\bf A}. The vector ${\bf m}\left({\bf r}\right)$ is the skyrmion texture as parametrized in \Eq{eq: skyrmion_def} and we set the helicity $\gamma = 0$ since it can be unitarily removed from the Hamiltonian.\cite{Loss_Majorana_skyrmion}  Hereafter, the exchange interaction $J$ includes the saturation magnetization $M$ of the magnet. We emphasize that the strengths of orbital effects and exchange interactions can be tuned independently. Experimentally, the coupling $J$ can be reduced by an insulating layer between the magnet and the superconductor.

The BdG Hamiltonian \Eq{eq: BdG_Skyrmion_full} has a generalized rotation symmetry and total angular momentum operator which, in absence of any vortices, reads $J_z = L_z + \frac{1}{2}\hbar\sigma_z$ where the orbital angular momentum reads $L_z = -i \hbar\partial_\theta$. The eigenvalues of $J_z$ provide a quantum number $m_J$ (in units of $\hbar$) labeling independent blocks of the BdG Hamiltonian. Angular momenta are half-odd-integer as required by the singlevaluedness of the wavefunction. We discretize the Hamiltonian according to $r \to r_j = ja$ with lattice constant $a$ chosen as the length unit ($a \equiv 1$) and the hopping parameter $t = \hbar^2/(2ma^2)$ is chosen as the energy unit ($t \equiv 1$, see SM~\ref{app: RTB} and the Methods section of Ref.~\onlinecite{Garnier_SkTopoSC} for additional technical details). For completeness, the lengthy expression of the discretized Hamiltonian $H_{m_J}^{(1)}$ corresponding to \Eq{eq: BdG_Skyrmion_full} is given in SM~\ref{app: RTB_f}. In all computations we use hard wall boundary conditions at the skyrmion's edge.

\subsection{Orbital effects without exchange and vortices}
Neglecting the exchange interaction ($J = 0$ in \Eq{eq: BdG_Skyrmion_full}), the relevant angular momentum operator is $J_z = L_z$. The discretized $m_L$-dependent Hamiltonian $H_{m_L}^{(2)}$ is given in SM~\ref{app: RTB_o}, with $m_L \in \mathbb{Z}$ the eigenvalue of $L_z$ (in units of $\hbar$). The vector potential contributes two terms: (1) a space-dependent renormalization of the chemical potential \via the ${\bf A}^2$ term, and (2) a term $\propto \mathds{1}$ that depends on both space and angular momentum.

\Fig{fig: spec_orbonly} below shows the numerically obtained BdG spectum contrasting the cases of weak and strong orbital effects, \ie $\nu =1$ (\Fig{fig: spec_orbonly}a) and $\nu = 5$ (\Fig{fig: spec_orbonly}b). In the latter case, we neglect the predicted appearance of vortices (see \Fig{fig: PhDiag}) to better isolate the orbital effects. The electronic local density of states (LDoS) along the $r = 0 \to r = \Rsk$ radial line in the $\nu =5$ case is shown in \Fig{fig: orbexch}a.
\begin{figure}[h!]
%\hspace*{-1.5cm}
\centering
\includegraphics[width=\columnwidth]{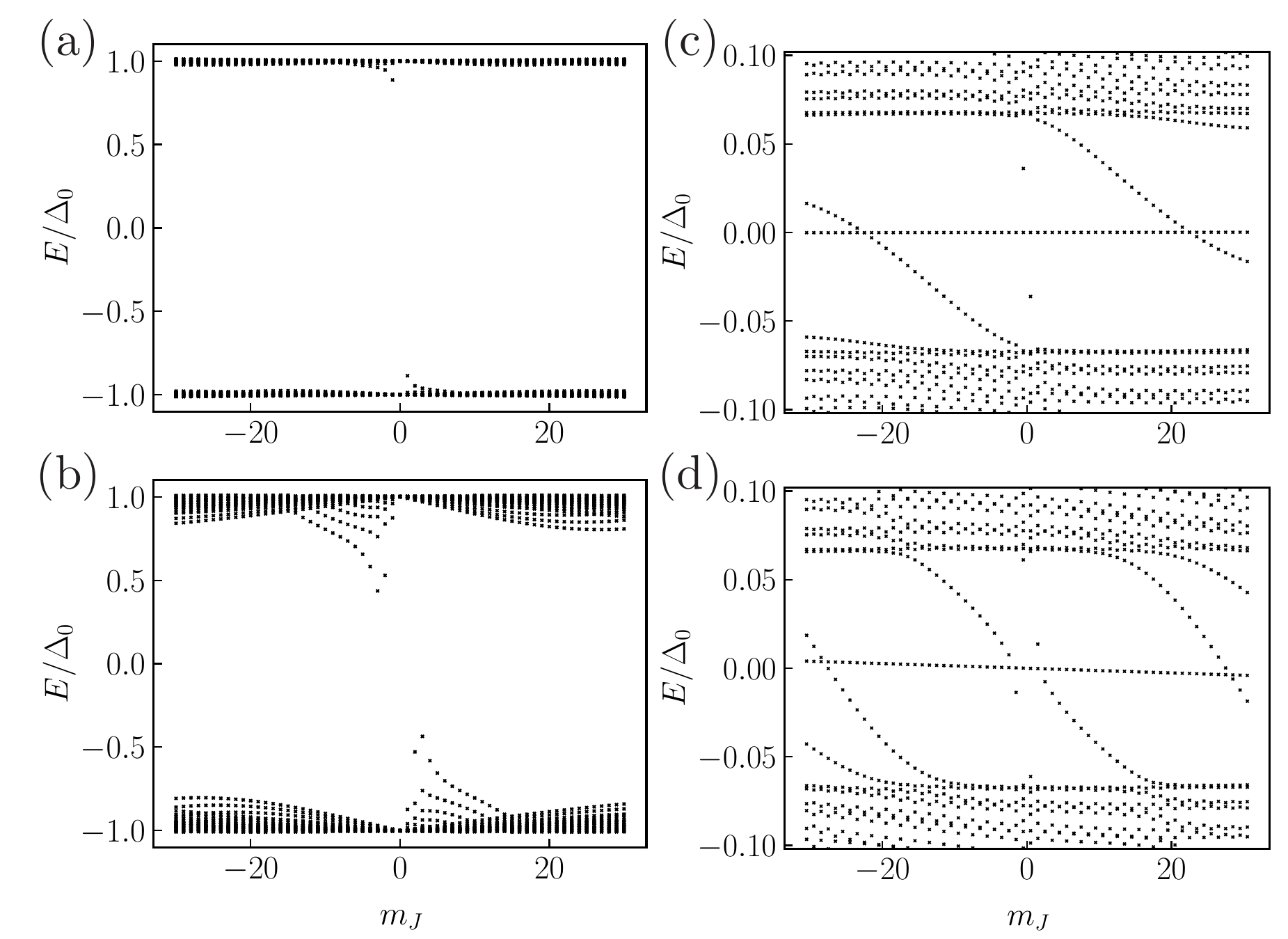}
\caption{Bogoliubov-de-Gennes spectra in absence of vortices. In absence of exchange coupling ($J=0$), the purely orbital effects are chosen weaker, $\nu = 1$, in (a) and stronger, $\nu=5$, in (b). In (c,d) topological superconductivity is caused by exchange coupling, $J/t = 0.2$, both in absence of orbital effects, $\nu=0$ in (c), and also in presence of orbital effects, $\nu=1$ in (d). The system parameters for all panels are $p = 10$, $L/a = 1000$, $\Do/t = 0.1$, $\mu/t = 0$ (see SM~\ref{app: RTB_o}).}
\label{fig: spec_orbonly}%
\end{figure}
\begin{figure}[h!]
\centering
\includegraphics[width=\columnwidth]{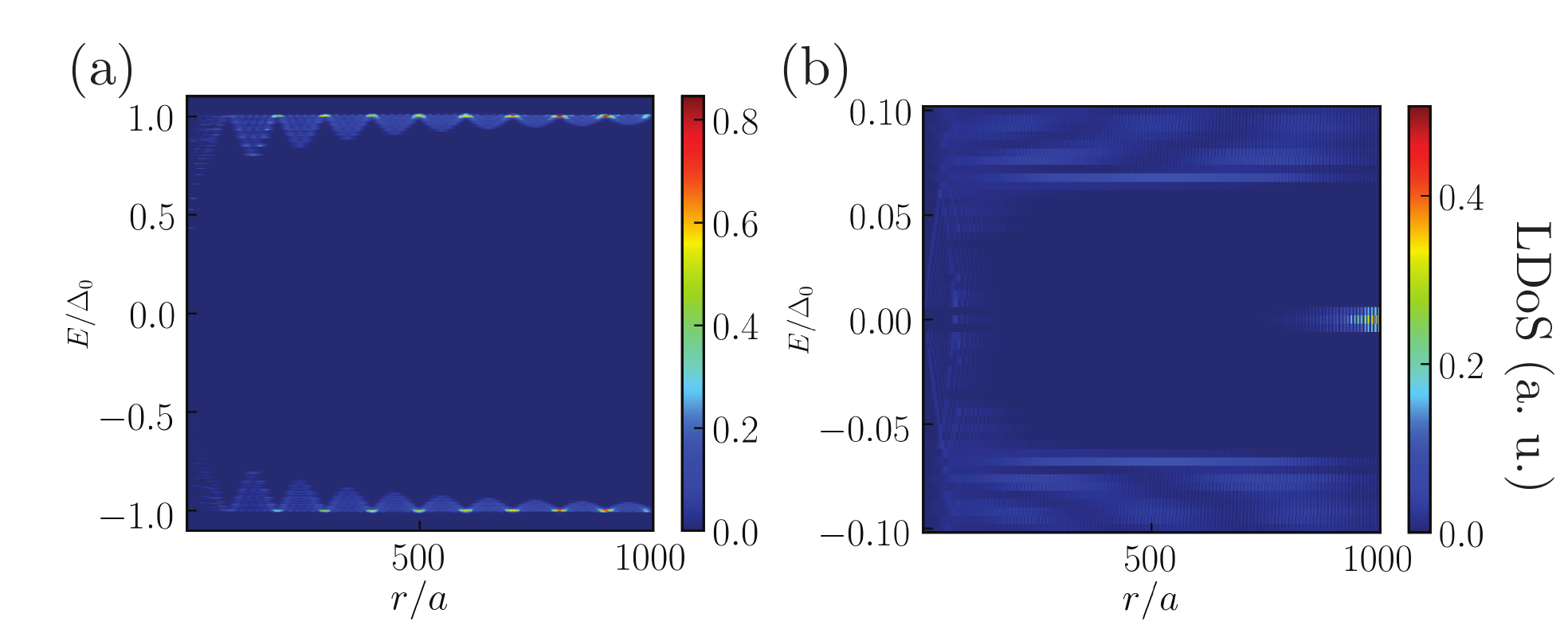}
\caption{Local density of states in absence of vortices, contrasting (a) purely orbital effects, $J= 0$ and $\nu = 5$ with (b) both exchange, $J/t = 0.2$, and orbital effects, $\nu = 1$. The models in (a), (b) are the same as in \Fig{fig: spec_orbonly}(b), (d), respectively. In both panels the system parameters are $p = 10$, $L/a = 1000$, $\Do/t = 0.1$, $\mu/t = 0$ (see SM~\ref{app: RTB_o}).}
\label{fig: orbexch}%
\end{figure}

We observe that the bare superconducting gap $\Do$ is preserved when orbital effects are added. This is somewhat expected since spin degeneracy is not lifted in the Hamiltonian. In the $\nu = 5$ case, a few localized states come down in energy but stick to the top (resp. bottom) of the gap for $m_L < 0$ (resp. $m_L >0$). We find that these states are largely localized (not necessarily near the core of the skyrmion) resembling an analogue of Caroli-de-Gennes-Matricon states.\cite{CdGM:1964} In fact, similar in-gap spectrum due to purely orbital effects was found in a self-consistent calculation\cite{Tanaka1993} of the magnetic field due to a preformed normal metal dot in a superconductor (without vortex).

\subsection{Orbital effects and exchange interaction}
When restoring the exchange interaction, the 2D BdG Hamiltonian in \Eq{eq: BdG_Skyrmion_full} has total angular momentum $J_z = L_z + \frac{1}{2}\hbar\sigma_z$ and the full radial tight-binding Hamiltonian $H_{m_J}^{(1)}$ is given in SM~\ref{app: RTB_f}.

To assess the consequences of orbital effects, we first return to the model with only exchange interaction ($\nu = 0$), which leads to topological superconductivity,\cite{Loss_Majorana_skyrmion, KovalevPRB2018, Morr_engineering, Garnier_SkTopoSC, Rex_SK_Vortex_2019} see \Fig{fig: spec_orbonly}c. For completess, let us recall briefly the properties of this topological superconductor. First, the effective gap is of $p$-wave origin with an amplitude given by $\Delta_{\rm eff} = \frac{\pi}{\lambda_s} \frac{\Delta_0}{J} \sqrt{J + \mu} $ where all energies are in units of $t$ and $\lambda_s$ is expressed in units of $a$. This evaluates to $\Delta_{\rm eff} \approx 7\% \Do$ which is consistent with the numerical data. Within the gap, there are two types of states: namely a nearly flat band located at the edge of the skyrmion together with dispersing states located near the core. These states are attributed to impurity-like states induced by the discretized magnetic texture. In the absence of orbital effects, the nearly-flat band is in fact slightly chiral and can be assigned a topological character thereby forming a chiral Majorana edge mode.\cite{Garnier_SkTopoSC, Rex_SK_Vortex_2019}

We now turn to the case of combined exchange and orbital effects for $\nu = 1$. \Fig{fig: orbexch}b shows that the topological superconductor described above essentially survives the inclusion of orbital effects. Specifically, the strongest effect on the spectrum is for the dispersing in-gap ``impurity states''. In contrast, the momentum $m_J^*$ at which the bulk gap closes still matches the value predicted for purely exchange coupling (see Ref.~\onlinecite{Garnier_SkTopoSC} for details).
Regarding the chiral edge mode, interestingly its velocity increases and changes sign. This can be phenomenologically explained by the chiral symmetry interpretation of Ref.~\onlinecite{Garnier_SkTopoSC}. In absence of orbital effects, the smallness of the velocity was attributed to an only weakly broken chiral symmetry given by  ${\cal S} = \sigma_y  \tau_y$. When including orbital effects ($\nu \neq 0$), different chiral-symmetry-breaking (CSB) terms appear, \eg the one proportional to $\id$. This term spatially decays as $r^{-1}$, so it should affect chirality stronger than the CSB term in the pure exchange model, which decays as $r^{-2}$.

\subsection{Adding superconducting vortices}
The interplay between exchange effects and vortices in a skyrmion proximitized by an $s$-wave superconductor was studied in Ref.~\onlinecite{Rex_SK_Vortex_2019} in absence of orbital effects. Here we analyse the full problem by including orbital effects together with the exchange interaction and a superconducting vortex.

Using the notation of the previous section, a superconducting vortex of winding number $\alpha$ is represented by modifying the pairing Hamiltonian $\H_{\rm SC} = \Do \tau_x$ in \Eq{eq: BdG_Skyrmion_full} to
\begin{align}
\H_{\rm v} = \Delta_\alpha\left(r\right) \, e^{i\alpha \theta} \tau_+ + \Delta_\alpha\left(r\right) \, e^{-i\alpha \theta} \tau_- 
\label{eq: vortex_param}
\end{align}
where $\tau_\pm = \left(\tau_x \pm i \tau_y\right)/2$.\footnote{We made sure of the consistency between the GL approach and the BdG one by requiring that the superconducting order parameter is an expectation value of the form \unexpanded{$\left\langle\psi \psi\right\rangle$}, which ensures that $\alpha$ has the same meaning in both approaches.} We neglect the spatial variation of $\Delta_\alpha\left(r\right)$ and assume that the amplitude of the order parameter is constant, independent of $\alpha$ and equal to $\Do$. In this situation, the total angular momentum reads $J_z = L_z + \frac{1}{2}\hbar\sigma_z - \frac{\alpha}{2}\hbar\tau_z$ and we still denote its eigenvalue by $m_J$. The momentum is quantized according to $m_J \in \mathbb{Z}$ (resp. $m_J \in \mathbb{Z} + \frac{1}{2}$) if $\alpha$ odd (resp. $\alpha$ even). The total radial Hamiltonian $H_{m_J}^{(3)}$ is given in SM~\ref{app: RTB_v}.

\begin{figure}[h!]
\centering
\includegraphics[width=\columnwidth]{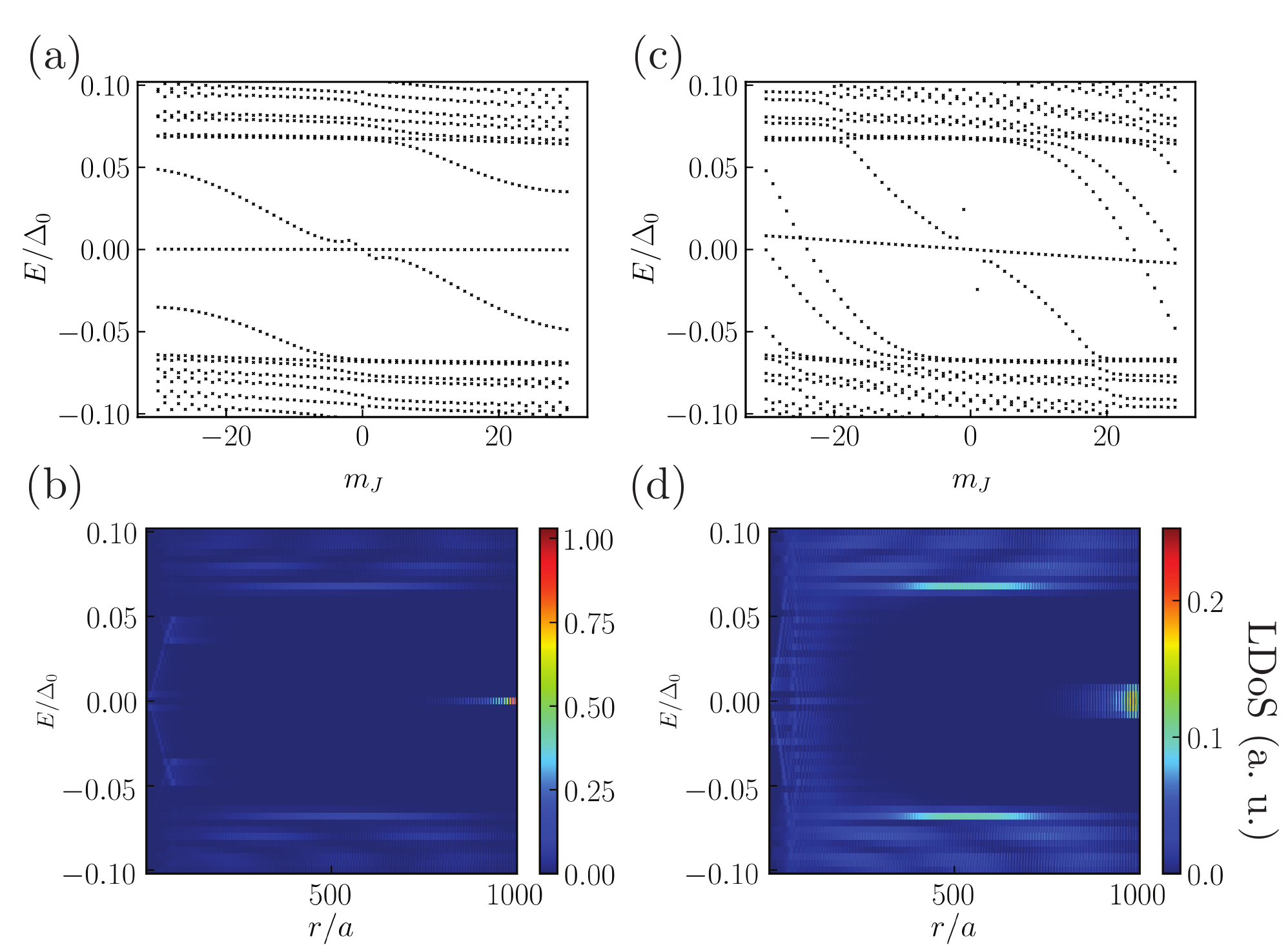}
\caption{BdG spectrum and LDoS in presence of superconducting vortex with winding number $\alpha = -1$. (a), (b) Spectrum and LDoS for $\nu = 0$ (no orbital effects). (c), (d) Spectrum and LDoS for $\nu = 2$. Fixed system parameters are $p = 10$, $L/a = 1000$, $J/t = 0.2$, $\Do/t = 0.1$, $\mu/t = 0$ (see SM~\ref{app: RTB_v}).}
\label{fig: vortex_results}%
\end{figure}

\Fig{fig: vortex_results}a, b show the excitation spectrum and the electronic LDoS for a vortex with $\alpha = -1$ without orbital effects, while \Fig{fig: vortex_results}c, d show the same with orbital effects of $\nu = 2$ (\cf the phase diagram in \Fig{fig: PhDiag}).
\Fig{fig: vortex_results}a,b show that the features of the topological superconductor of the exchange model are almost unaffected by the presence of the vortex (without orbital effects), in accord with Ref.~\onlinecite{Rex_SK_Vortex_2019}.

\Fig{fig: vortex_results}c,d show that the inclusion of orbital coupling leads to similar effects as in the case of the absence of vortex (previous subsection), namely, (1) it mixes the in-gap impurity states (similar impurity states were found due to purely orbital effects in presence of vortex on a metallic dot in a superconductor\cite{Tanaka1993}), and (2) it changes the slope of the topological chiral mode.
For these specific parameters, note that the chiral symmetry breaking due to the vortex (proportional to $\alpha/r^2$, see \Eq{eq: TB_Ham_vortex} in SM~\ref{app: RTB_v}) adds constructively to the symmetry breaking by the orbital term (proportional to $\nu/r$, see \Eq{eq: TB_Ham_full_supp} in SM~\ref{app: RTB_f}), because the change in the slope of the chiral mode is significantly higher when the vortex is included together with the orbital terms.

\section{Discussion}

Our analysis shows that the exchange-induced topological phase is robust to the inclusion of skyrmion-generated orbital effects as well as superconducting vortices. These effects have to be taken into account since the magnetic field generated by an isolated skyrmion can reach the ${\rm mT}$ range\cite{Halbach_exp:2018}. We have shown using a Ginzburg-Landau approach that proximitizing a skyrmion with a superconductor does not necessarily lead to the formation of vortices. However, even if the superconductor does not develop vortices, the electrons still experience a magnetic vector potential whose effects on the topological superconducting phase were not fully understood in materials without spin-orbit coupling. Our results demonstrate that the inclusion of orbital effects does not invalidate the previously established understanding of the topological superconductor and contributes to making the skyrmion-superconductor hybrid structure a promising platform for the realization of topological superconductivity. Note however that we have neglected the Zeeman effect that would effectively render the exchange interaction anisotropic (see \Eq{eq: blat}). Nevertheless  this renormalization is far below the bare exchange interaction strength, so would not matter even if the exchange strength was reduced by a non-magnetic insulating layer between the skyrmion and the superconductor. 

Because isolated skyrmions with arbitrary winding numbers and helicity  are theoretically more likely to induce topological superconductivity,\cite{Loss_Majorana_skyrmion,Garnier_SkTopoSC} a possible interesting direction would be 
to extend the present magnetostatic calculations for such skyrmions.
More generally, our work calls for a fully self-consistent calculation of both the magnetic and the superconducting order to confirm all the features of the system.

\section{Acknowledgments}
The authors acknowledge useful conversation with Freek Massee, Marco Aprili, Stanislas Rohart and Vardan Kaladzhyan. M. G. thanks Marc Gabay for stimulating and insightful discussions.

\bibliography{orbital_paper}

\cleardoublepage
\onecolumngrid
\appendix
\renewcommand\appendixname{Supplementary Material}
\renewcommand{\figurename}{Fig. \thesection}
\setcounter{figure}{0}

\begin{center}
{\bf \large Supplementary Material: Magnetic-skyrmion-induced orbital effects in superconductors} \\[0.5cm]
Maxime Garnier, Andrej Mesaros and Pascal Simon  \\[0.5cm]
\textit{Laboratoire de Physique des Solides, UMR 8502, CNRS, \\
Universit\'{e} Paris-Sud, Universit\'{e} Paris-Saclay, 91405 Orsay, France}
\end{center}

\section{Magnetostatics of the skyrmion lattice}
\label{sec: app_SkL}
\setcounter{figure}{0}
In this appendix we support our hypothesis that the perpendicular induction $B_z$ created by a single skyrmion is proportional to the $z$ component of the magnetization. We do so by focusing on a skyrmion lattice following Ref.~\onlinecite{QinWang_SKL_2018}.
A triangular skyrmion lattice can be approximated using a superposition of three helical spin orders with wavevectors ${\bf Q}_{i=1,\, 2,\, 3}$ yielding the so-called triple-${\bf Q}$ parametrization\cite{TokNag2013,SK_Chiral_Mag_Wiesendanger}. In three-dimensional space equiped with the unit vectors ${\bf u}_{x,y,z}$, consider an infinitely thin magnetic film that lies in the $z = 0$ plane. The three wavevectors ${\bf Q}_i$ have the same norm $\left\vert{\bf Q}_{i}\right\vert = Q$ for all $i$ and they make a $2\pi/3$ angle with respect to each other. Our precise choice is ${\bf Q}_1 = Q {\bf u}_x$, ${\bf Q}_2 = Q \left(-{\bf u}_x + \sqrt{3}{\bf u}_y\right)/2$ and ${\bf Q}_3 = Q \left(-{\bf u}_x - \sqrt{3}{\bf u}_y\right)/2$.
%so that $\sum_i {\bf Q}_{i} = {\bf 0}$.
For a \neel skyrmion lattice, the magnetization ${\bf m}\left({\bf r}, z\right)$ with ${\bf r} = \left(x, y\right)$ reads
\begin{align}
\begin{split}
& {\bf m}_{\rm lat}\left({\bf r}, z \right)  = m_0 \, \delta\left( z \right) {\bf u}_z \\  & + A \sum_{i = 1}^3 \left[ \cos\left({\bf Q}_i\cdot{\bf r}\right) {\bf u}_z + \sin\left({\bf Q}_i\cdot{\bf r}\right){\bf u}_i  \right] \, \delta \left( z \right)
\end{split}
\label{eq: SkL_approx}
\end{align}
where we have defined the unit vectors ${\bf u}_i = {\bf Q}_{i}/Q$. We also define the skyrmion radius $\Rsk$ as $\Rsk = 2\pi/\left(Q\sqrt{3}\right)$ (see caption of Fig. A \ref{fig: SkL_plot}). \Eq{eq: SkL_approx} is an approximation in the sense that it is not a proper micromagnetic solution since it is not normalized.

In terms of the thickness $h$ and the saturation magnetization $M_{\rm lat}$ of the magnetic film, the parameter $A$ reads $A= M_{\rm lat} h$. In this approach, we define the skyrmion radius $\Rsk$ as $\Rsk = 2\pi/\left(Q\sqrt{3}\right)$. An example of such a \neel skyrmion lattice is presented in Fig. A \ref{fig: SkL_plot}.
\begin{figure}[h!]
\centering
\includegraphics[scale=0.6]{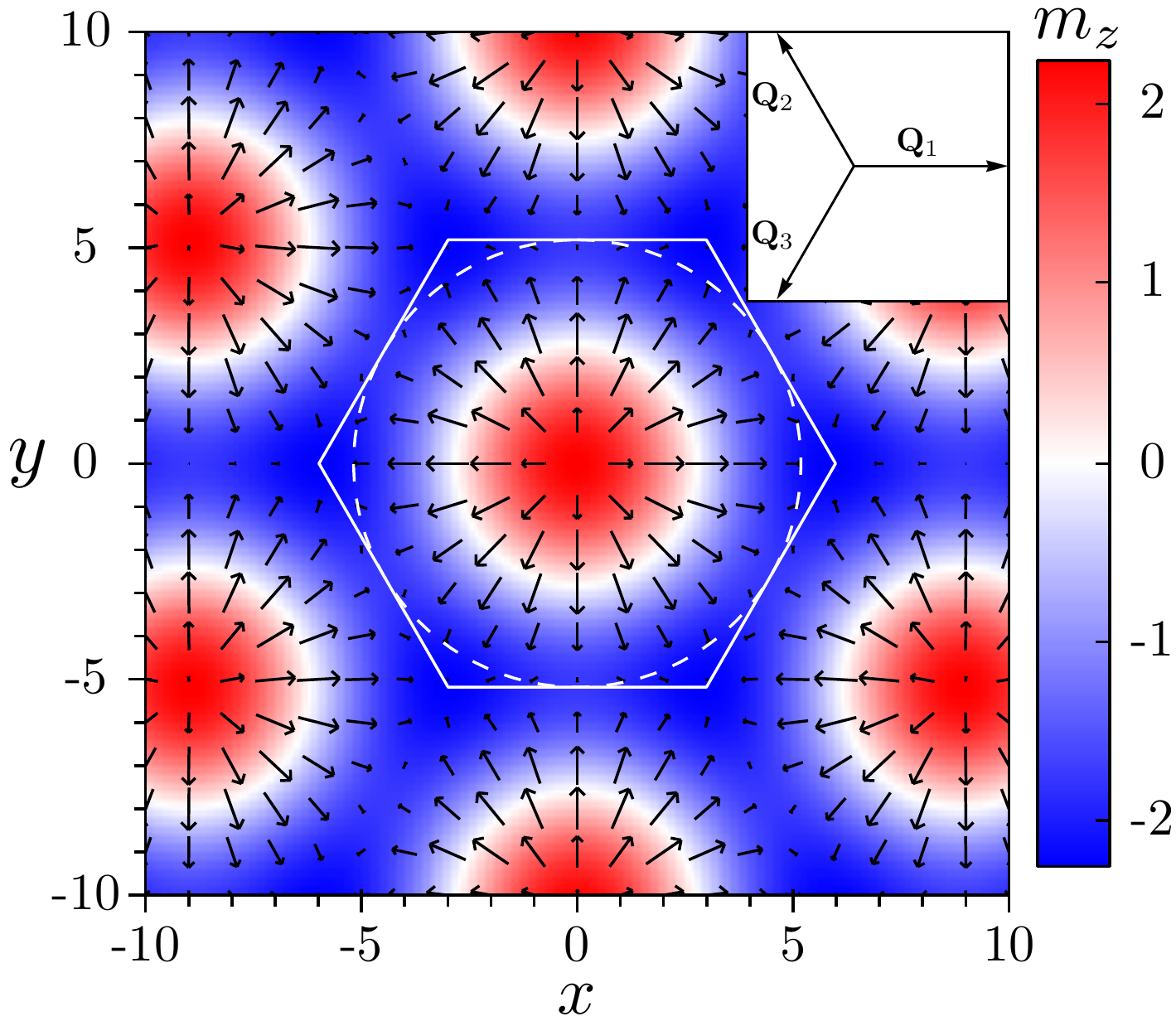} %width=\columnwidth %scale = .5
\caption{Triangular \neel skyrmion lattice as obtained from \Eq{eq: SkL_approx}. The black arrows represent the in-plane components of the magnetization $\left(m_{\rm lat}^x, m_{\rm lat}^y\right)$ while the colormap represents the out-of-plane component $m_{\rm lat}^z$. We define the skyrmion radius $\Rsk$ as the radius  $\Rsk = 2\pi/\left(Q\sqrt{3}\right)$ of the white dashed circle inscribed in the unit cell (white hexagon). The parameters used are $A=1$, $m_0 = -0.755$ and $Q = 0.7$.}
\label{fig: SkL_plot}
\end{figure}
One way to compute the magnetic field\footnote{Strictly speaking ${\bf  B}$ is the magnetic induction but as we focus on the region outside the magnetized medium, the magnetic induction ${\bf  B}$ and the magnetic field ${\bf  H}$ are related by ${\bf  B} = \mu_0 {\bf  H}$ so that we call ${\bf  B}$ the magnetic field.} ${\bf  B}_{\rm lat} \left({\bf r}, z \right)$ created by the magnetization distribution is to compute the magnetic vector potential ${\bf A}_{\rm lat}\left({\bf r}, z \right)$ since ${\bf  B}_{\rm lat} = {\bm \nabla} \cross {\bf  A}_{\rm lat}$. Then, ${\bf A}_{\rm lat}$ is found by solving the Poisson equation $\nabla^2{\bf A}_{\rm lat}\left({\bf r}, z \right) = -\mu_0 {\vb J}_m\left({\bf r}, z \right) $ in the Coulomb gauge ${\bm \nabla} \cdot {\bf A}_{\rm lat} = 0$, where ${\bf J}_m\left({\bf r}, z \right) =  {\bm \nabla} \cross {\bf  m}_{\rm lat} \left({\bf r}, z \right)$ is the Amperean current density that contains both ``bulk'' and ``surface'' contributions. Taking the curl of the solution ${\bf A}_{\rm lat}\left({\bf r}, z \right)$ of this equation yields %\cite{Mallinson:1973}
\begin{align}
\begin{split}
{\bf  B}_{\rm lat} \left({\bf r}, z \right) & = \mu_0 \frac{A}{2} e^{-Q\left\vert z \right\vert} \sum_{i = 1}^3 \left[ Q \left(1 - \, {\rm sgn} \left( z\right) \right)\cos\left({\bf Q}_i\cdot{\bf r}\right){\bf u}_z \right.\\& \left.+ \left(Q \, {\rm sgn} \left( z\right) -Q + 2\delta \left( z \right)\right)\sin\left({\bf Q}_i\cdot{\bf r}\right) {\bf u}_i   \right]
\end{split}
\label{eq: blat}
\end{align}
This field displays the ``single-sided flux'' phenomenon or Halbach effect\cite{QinWang_SKL_2018,Halbach_exp:2018,Mallinson:1973} meaning that the magnetic field is only present on one side of the plane. The apparent discontinuity in the perpendicular component of the magnetic induction is an artifact of the model and can be regularized by taking into account the finite thickness of the magnetic film so that this effect is indeed physical\cite{Halbach_exp:2018,Mallinson:1973}. The perpendicular decay length of the magnetic field is given by $Q^{-1}$ and can be expressed in terms of the skyrmion radius $\Rsk$ as $Q^{-1} = \sqrt{3}\Rsk/\left(2 \pi\right) \approx 0.3 \, \Rsk$.

With the idea to proximitize the skyrmion lattice by a superconductor, we are only interested in the magnetic field near the plane \ie $\left\vert Q z \right\vert \to 0$. In the case discussed here, the relevant limit is $Q z \to 0^-$.
In this limit, the $z$ component of the magnetic field reads
\begin{align}
B^z_{\rm lat}\left({\bf r}, 0^- \right) & =    \mu_0 A Q \sum_{i = 1}^3   \cos\left({\bf Q}_i\cdot{\bf r}\right)
\label{eq: Bfield_planelim}
\end{align}
\ie close to the plane $B^z_{\rm lat}$ is proportional to the $z$ component of the magnetization. Focusing on $z \ll Q^{-1}$,  we argue that this result applies to our case of the isolated skyrmion as long as the thickness $d$ of the superconductor is much smaller than the skyrmion radius since $\Rsk \propto Q^{-1}$. Additionally, we have $A = M_{\rm lat} h$, we find that the amplitude of the magnetic field near the surface of the magnet is given by $\mu_0 M_{\rm lat} Q h$. Note that this result also holds in the case of a Bloch skyrmion lattice with the only difference that the magnetic field is evenly shared between the two sides of the plane\cite{QinWang_SKL_2018}. 

\section{Influence of parameters on the phase diagram}
\label{sec: app_PhD}
\setcounter{figure}{0}
The phase diagram displayed in \Fig{fig: PhDiag} in the main text was computed for a $p = 4$ skyrmion with a cutoff $k_s l = 10^{-1}$ and we neglected the magnetic energy term proportional to $\mu_0 M/B_c$ where $B_c$ is the thermodynamic critical field of the superconductor. We now discuss their effects.

\subsection{Effect of the short-distance cutoff}
\label{app: cutoff}
\begin{figure}[h!]
\centering
\includegraphics[scale = .4]{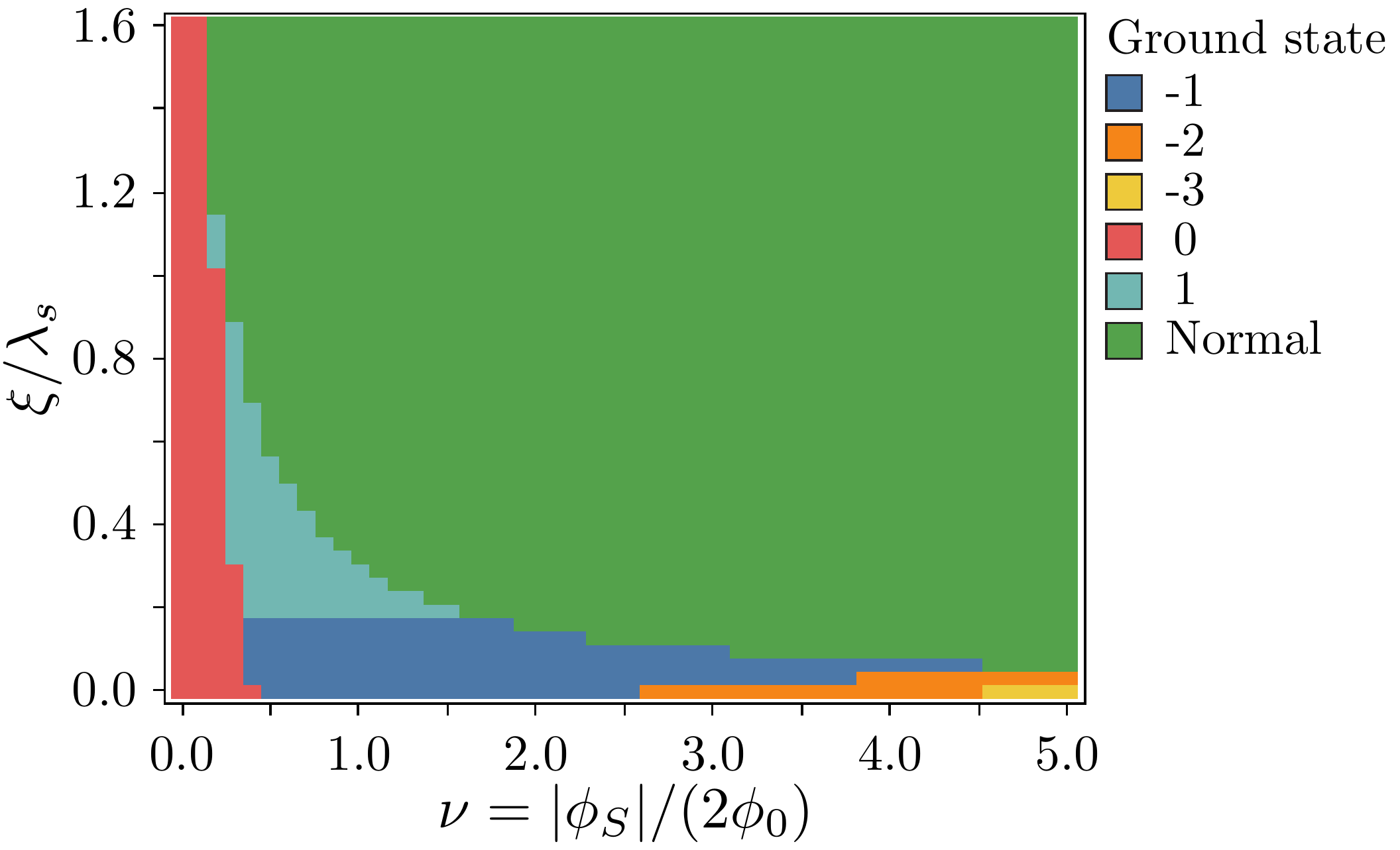} %width=\columnwidth
\caption{Phase diagram for the same parameters as in \Fig{fig: PhDiag} but with a smaller cutoff $k_s l = 10^{-12}$.}
\label{fig: PhDiag_cutoff}
\end{figure}

We expect that reducing the cutoff will make the superconducting state with winding $\alpha = 0$ energetically defavorable since is will give more weight to the divergence as $r \to 0$. This is indeed what we obtain in Fig.~B~\ref{fig: PhDiag_cutoff} by changing $k_s l = 10^{-1}$ to $k_s l = 10^{-12}$.
The normal-superconductor transition line is almost unaffected by this change but the $\alpha = 0$ phase still exists for weak enough skyrmions.

\subsection{Effect of the magnetic energy term}
\label{app: magterm}
\begin{figure}[h!]
\centering
\includegraphics[scale = .4]{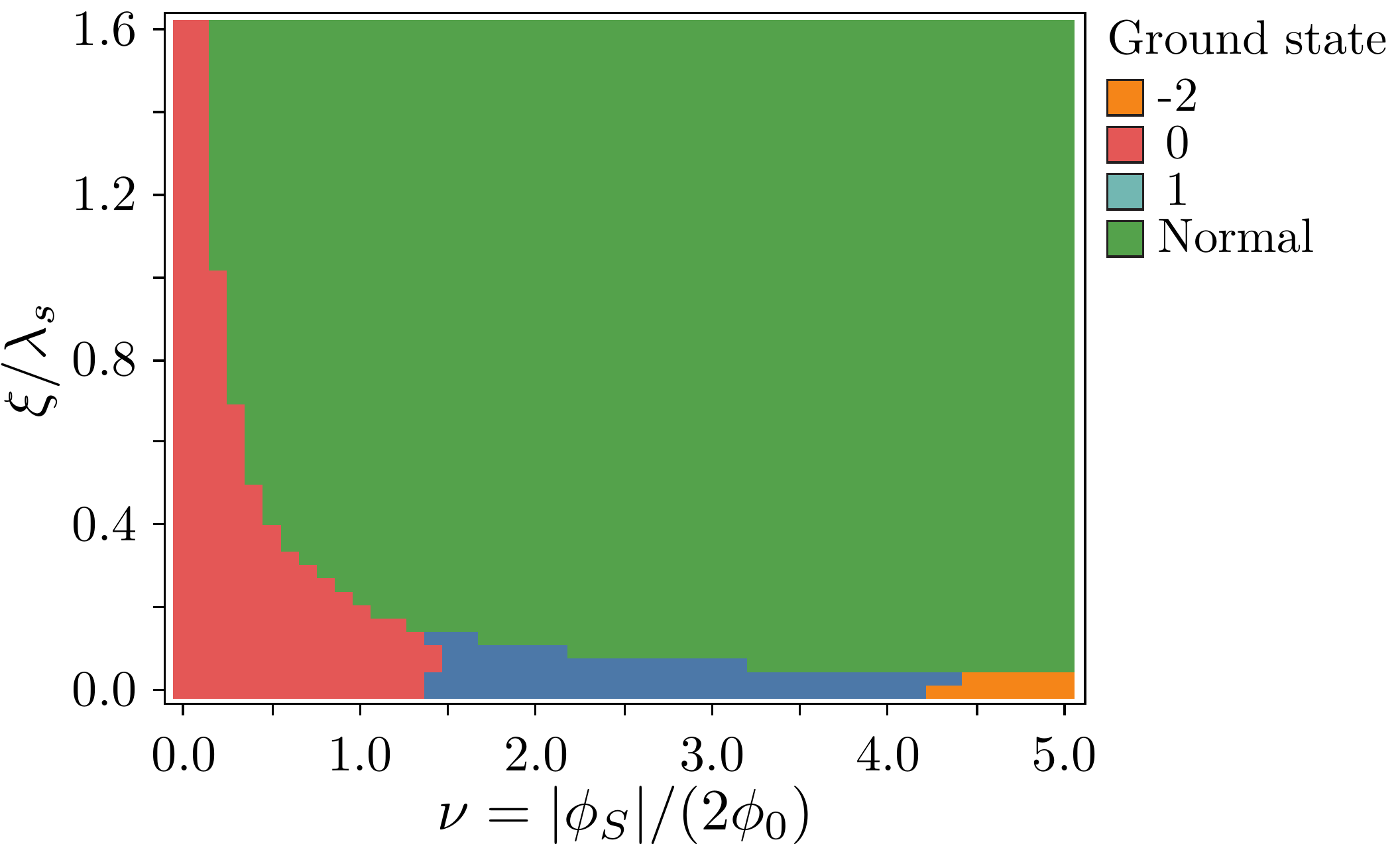} 
\caption{Phase diagram for the same parameters as in \Fig{fig: PhDiag} but the purely magnetic term is included with the approximation $\mu_0 M/ B_c = 1$ for all $\nu$.}
\label{fig: PhDiagBc}
\end{figure}

We now reinstate the magnetic energy term we have neglected so far. As it is $\alpha$-independent, it's only effect will be to move the normal-superconducting transition line depending on the ratio $\mu_0 M/B_c$. However, in our parametrization the parameter $\nu$ is not independent from $M$. In order to estimate the effects of this term, we set $\mu_0 M/B_c = 1$ for all $\nu$. This is a crude approximation that largely overestimates the strength of this term on a large area of the phase diagram. The phase diagram is shown in Fig.~B~\ref{fig: PhDiagBc} 

This shows that the inclusion of the magnetic term does not qualitatively affect our results. It further shows that the inhomogeneity of the skyrmion yields a stable superconducting phase even if $\mu_0 M > B_c$.

\subsection{Influence of the skyrmion size}
\label{app: SkSize}
\begin{figure}[h!]
\centering
%\hspace*{-.5cm}
\includegraphics[scale = 1]{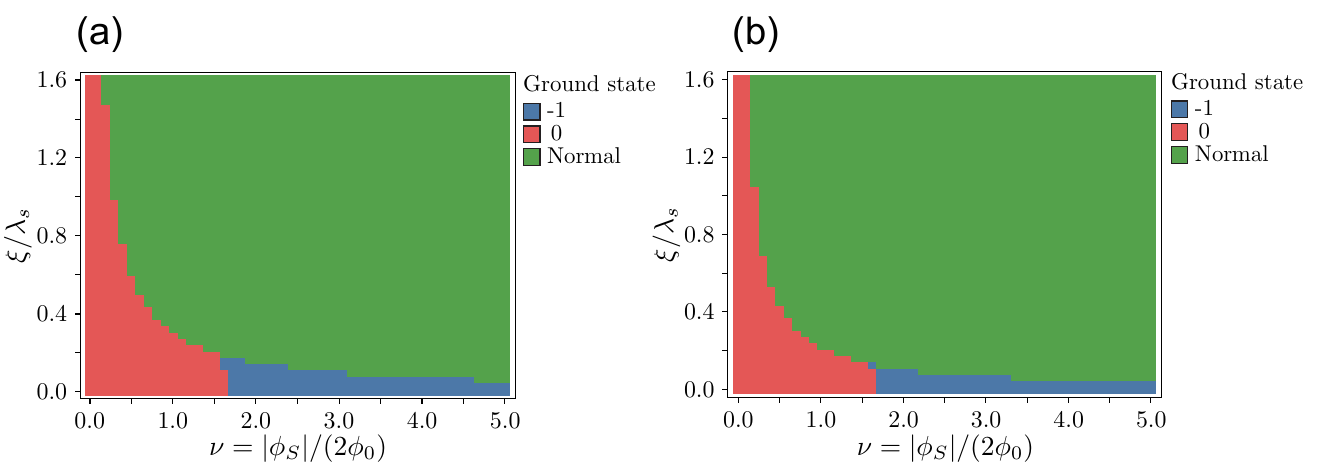}
\caption{Phase diagram of a $p = 10$ skyrmion with cutoff $k_s l = 10^{-1}$. (a) without the magnetic energy term and (b) with the magnetic energy term and $\mu_0 M/ B_c = 1$  for all $\nu$.}
\label{fig: PhDiagSkSize}
\end{figure}

The results for a $p = 10$ skyrmion are displayed in Fig.~B~\ref{fig: PhDiagSkSize}. These results show that the conclusions of the main text hold qualitatively except that winding number different from -1 have now disappeared from the phase diagram.

\section{Radial tight-binding Hamiltonians}
\label{app: RTB}
\setcounter{figure}{0}
In this section we give the expression of the second-quantized discretized radial Hamiltonians used to obtain the results of the main text. In all Hamiltonians we have introduced $t = \hbar^2/\left(2m a^2\right)$, ${\rm k}_s = k_s a$ where $a \equiv 1$ is the radial lattice spacing chosen as the distance unit. We use the notation ${\cal C}^\dagger_{j} =  \left(c^\dagger_\up \left(j a\right), c^\dagger_\down \left(j a\right),c_\down \left(j a\right), -c_\up \left(j a\right) \right)$ and $c_\sigma\left(j a\right)$ to denote the discretized versions of the spinor $\Psi\left({\bf r}\right)$ and the field operator $\psi_\sigma\left({\bf r}\right)$. We have chosen $j = 0.5, \dots, L + 0.5$ to avoid the singularity at the origin.

\subsection{Orbital effects and exchange interaction}
\label{app: RTB_f}
This case corresponds to \Eq{eq: BdG_Skyrmion_full} of the main text. The total angular momentum is $J_z = L_z + \frac{1}{2}\hbar\sigma_z$ with eigenvalue $m_J$ (in units of $\hbar$). The discretized Hamiltonian reads
\begin{align}
\begin{split}
H_{m_J}^{(1)}  &= \sum_{j} -t \, {\cal C}^\dagger_{j+1} \tau_z {\cal C}_j  + \hc   + {\cal C}^\dagger_j \left[2t - \mu - \frac{t}{4 j^2}\left(1 - 4 m_J^2 - q^2 \right) - \frac{t}{j^2}q m_J \sigma_z   +\frac{t}{4} {\rm k}_s^2 \nu^2 \left(\frac{\cos\left({\rm k}_s j\right)}{{\rm k}_s j} + \sin\left({\rm k}_s j\right)\right)^2\right]\tau_z {\cal C}_j  \\ & + {\cal C}^\dagger_j \left[t \frac{{\rm k}_s \nu}{j}\left(\frac{\cos\left({\rm k}_s j\right)}{{\rm k}_s j} + \sin\left({\rm k}_s j\right)\right)\left(m_J \id - \frac{q}{2}\sigma_z\right) + J \cos\left({\rm k}_s j\right) \sigma_z + J \sin\left({\rm k}_s j\right) \sigma_x + \Delta_0 \tau_x\right] {\cal C}_j
\end{split}
\label{eq: TB_Ham_full_supp}
\end{align}

\subsection{Orbital effects without exchange and vortices}
\label{app: RTB_o}
This case corresponds to setting $J = 0$ in \Eq{eq: BdG_Skyrmion_full} of the main text. or equivalently $J = q = 0$ in \Eq{eq: TB_Ham_full_supp}. The total angular momentum is $J_z = L_z$ with eigenvalue $m_L$ (in units of $\hbar$). The discretized Hamiltonian is
\begin{align}
\begin{split}
H_{m_L}^{(2)} = \sum_{j} -t \, {\cal C}^\dagger_{j+1} \tau_z {\cal C}_j & + \hc   + {\cal C}^\dagger_j \left[2t - \mu - \frac{t}{4 j^2}\left(1 - 4 m_L^2\right) +\frac{t}{4} {\rm k}_s^2 \nu^2 \left(\frac{\cos\left({\rm k}_s j\right)}{{\rm k}_s j} + \sin\left({\rm k}_s j\right)\right)^2\right]\tau_z {\cal C}_j  \\ & + {\cal C}^\dagger_j \left[t \frac{{\rm k}_s \nu}{j}\left(\frac{\cos\left({\rm k}_s j\right)}{{\rm k}_s j} + \sin\left({\rm k}_s j\right)\right)m_L \id + \Delta_0 \tau_x\right] {\cal C}_j
\end{split}
\label{eq: TB_Ham_orb}
\end{align}

\subsection{Orbital effects, exchange interaction and superconducting vortex}
\label{app: RTB_v}
Replacing the superconducting pairing term in \Eq{eq: BdG_Skyrmion_full}  by \Eq{eq: vortex_param} (both in the main text), the total angular momentum is $J_z = L_z + \frac{1}{2}\hbar\sigma_z - \frac{\alpha}{2}\hbar\tau_z$ with eigenvalue $m_J$ (in units of $\hbar$) where $\alpha$ is the vortex winding. The discretized Hamiltonian reads
\begin{align}
\begin{split}
H_{m_J}^{(3)}  &= \sum_{j = 1}^L -t \, {\cal C}^\dagger_{j+1} \tau_z {\cal C}_j  + \hc   + {\cal C}^\dagger_j \left[2t - \mu - \frac{t}{4 j^2}\left(1 - 4 m_J^2 - q^2 - \alpha^2\right)  +\frac{t}{4} {\rm k}_s^2 \nu^2 \left(\frac{\cos\left({\rm k}_s j\right)}{{\rm k}_s j} + \sin\left({\rm k}_s j\right)\right)^2\right]\tau_z {\cal C}_j  \\ & + {\cal C}^\dagger_j \left[t \frac{{\rm k}_s \nu}{j}\left(\frac{\cos\left({\rm k}_s j\right)}{{\rm k}_s j} + \sin\left({\rm k}_s j\right)\right)\left(m_J \id - \frac{q}{2}\sigma_z + \frac{\alpha}{2}\tau_z\right) + \frac{t}{j^2}\left(- q m_J \sigma_z + \alpha m_J \tau_z - \frac{\alpha q}{4} \sigma_z\tau_z\right)\tau_z
\right.\\ & \left. + J \cos\left({\rm k}_s j\right) \sigma_z + J \sin\left({\rm k}_s j\right) \sigma_x + \Delta_\alpha\left(r\right) \, \tau_x\right] {\cal C}_j
\end{split}
\label{eq: TB_Ham_vortex}
\end{align}

\end{document}